\documentclass[superscriptaddress,twocolumn,showpacs,prb]{revtex4-1}
\usepackage[utf8]{inputenc} 
\usepackage{amsmath}
\usepackage{braket}
\usepackage{graphicx}
\usepackage{amsfonts}
\usepackage{pgfplots}
\usepackage{csquotes}
\usepackage{hhline}
\usepackage{amssymb}
\usepackage{listings}
\usepackage{color}
\usepackage{subfigure}
\usepackage{graphicx}

\definecolor{codegreen}{rgb}{0,0.6,0}
\definecolor{codegray}{rgb}{0.5,0.5,0.5}
\definecolor{codepurple}{rgb}{0.58,0,0.82}
\definecolor{backcolour}{rgb}{0.95,0.95,0.92}
 
\lstdefinestyle{mystyle}{
    backgroundcolor=\color{backcolour},   
    commentstyle=\color{codegreen},
    keywordstyle=\color{magenta},
    numberstyle=\tiny\color{codegray},
    stringstyle=\color{codepurple},
    basicstyle=\footnotesize,
    breakatwhitespace=false,         
    breaklines=true,                 
    captionpos=b,                    
    keepspaces=true,                 
    numbers=left,                    
    numbersep=5pt,                  
    showspaces=false,                
    showstringspaces=false,
    showtabs=false,                  
    tabsize=2
}
 
\usepackage{subfigure}

\usepackage{dcolumn}
\usepackage{tabularx}
\usepackage[colorlinks=true,linkcolor=blue,citecolor=blue,urlcolor=blue]{hyperref}
\usepackage{longtable}
\usepackage{braket}
\usepackage{float}
\newcolumntype{C}{>{\centering\arraybackslash}X}
\begin{document}
\title{Quantum simulation of negative hydrogen ion using variational quantum eigensolver on IBM quantum computer}

\author{Shubham Kumar}
\email{shubhamkumar.kumar@gmail.com}
\affiliation{Department of Physics, Central University of Jharkhand, Brambe, Ranchi, India}
\author{Rahul Pratap Singh}
\email{singhprataprahul97@gmail.com}
\affiliation{Indian Institute of Science Education and Research Kolkata,\\ Mohanpur 741246, West Bengal, India}

\author{Bikash K. Behera}
\email{bkb18rs025@iiserkol.ac.in}
\author{Prasanta K. Panigrahi}
\email{pprasanta@iiserkol.ac.in}
\affiliation{Department of Physical Sciences,\\ Indian Institute of Science Education and Research Kolkata, Mohanpur 741246, West Bengal, India}

\begin{abstract}
The negative hydrogen ion is the first three body quantum problem whose ground state energy is theoretically calculated using the ``Chandrasekhar Wavefunction'' that accounts for the electron-electron correlation \cite{qH_ChandrasekharAPJ1944}. The best value of ground state energy is obtained by photodetachment experiment using lasers in the laboratory. Solving multi-body systems is a daunting task in quantum mechanics as it includes choosing a trial wavefunction and the calculation of integrals for the system that becomes almost impossible for systems with three or more particles. This difficulty can be addressed by quantum computers. They have emerged as a tool to address different electronic structure problems with remarkable efficiency. Here, we show the quantum simulation of $H^{-}$ ion to calculate it's ground state energy in IBM quantum computer. We observe that the quantum computer is efficient in preparing the correlated wavefunction of $H^{-}$ and shows it as a bound entity as the ground state energy is found to be lower than that of Hydrogen atom. We use a recently developed algorithm known as ``Variational Quantum Eigensolver'' \cite{qH_PeruzzoNatComm2014,qH_KandalaNat2017} and implement it in IBM's 5-qubit quantum chips ``ibmqx2" and ``ibmqx4". An optimization routine is performed on a classical computer by running quantum chemistry program and codes in QISKit to converge the energy to the minimum. We also present a comparison of different optimization routines and encoding methods used to converge the energy value to the minimum. The circuit is parametrized by 12 arbitrary angles and is thus used to create different trial wavefuntions by varying the parameters. The technique can be used to solve various many body problems \cite{qH_Wiesner1996} with great efficiency.   
\end{abstract}


\maketitle

\section{Introduction}

Quantum simulation \cite{qH_GeorgescuRMP2014} is a fast-growing field which promises to have profound applications in the field of condensed-matter physics, nuclear physics, quantum cosmology, quantum chemistry, and quantum biology. According to Feynman \cite{qH_Feynman1982}, a quantum computer can deal with exponential amount of information without using exponential amount of resources. He pointed out that physical systems can be studied on quantum computers. A decade later, Lloyd proved that a quantum computer can actually be used as a universal quantum simulator \cite{qH_LlyodSci1996}. Simulations of quantum systems have always been a tough job even for present generation supercomputers, due to the exponential explosion when the system size increases. Hence, the use of quantum computer for quantum simulation is essential for near term future applications. A wide range of problems in condensed-matter physics e.g., quantum phase transitions \cite{qH_SachdevBook1999}, quantum magnetism \cite{qH_SachadevNatPhys2008}, in quantum chemistry e.g., calculating molecular energy values
\cite{qH_LanyonNatChem2010}, in nuclear physics e.g., studying atomic nucleus dynamics \cite{qH_DumitrescuPRL2018}, in quantum biology e.g., analyzing the structure of protein and DNA \cite{qH_LamberNatPhys2013}, in quantum cosmology e.g., structuring space-time curves \cite{qH_LloydarXiv2018,qH_ZizziGRG2001} can be addressed using quantum computers. A detailed review and future applications and implications of quantum simulation can be found from these Refs. \cite{qH_GeorgescuRMP2014,qH_TrabesingerNatPhys2012,qH_CiracNatPhys2012,qH_SchaetzNJP2013}. Different architectures such as optical lattice- \cite{qH_GrossSci2017,qH_TarruellarXiv2018}, trapped ions- \cite{qH_MonroeNat2016}, nuclear spins- \cite{qH_KaneNat1998}, superconducting qubit- \cite{qH_DevoretSci2013} based quantum computers have been extensively used in the past for simulation of quantum systems. 

Since 2016, IBM provides the composer on its website which is a cloud-based quantum computing platform \cite{qH_IBMQ}. Any user can give a quantum circuit on the five-, and sixteen-qubit devices for a real run or simulation which is available with the help of QISKit Terra and use it by changing backend to perform a run or simulation. IBM Q Experience has now been used to perform a number of real experiments on the quantum chips. The real experiments include quantum simulation \cite{qH_KandalaNat2017,qH_HaldararXiv2018,qH_MalickRG2019,qH_4BKB6arXiv2018,qH_VishnuQIP2018,qH_SchuldEPL2017,qH_1BKB6arXiv2018,qH_2BKB6arXiv2018,qH_ManabputraarXiv2018,qH_ViyuelanpjQI2018}, developing quantum algorithms \cite{qH_GarciaJAMP2018,qH_RounakarXiv2018,qH_SisodiaQIP2017,qH_GangopadhyayQIP2018,qH_DeffnerHel2017,qH_YalcinkayaPRA2017,qH_5BKB6arXiv2018}, testing of quantum information theoretical tasks \cite{qH_VishnuQIP2018,qH_GarciaJAMP2018,qH_HuffmanPRA2017,qH_AlsinaPRA2016,qH_KalraarXiv2017}, quantum cryptography \cite{qH_BeheraQIP2017,qH_Plesa2018,qH_MajumderarXiv2017}, quantum error correction \cite{qH_GhoshQIP2018,qH_Roffe2018,qH_SatyajitQIP2018,qH_3BKB6arXiv2018}, quantum applications \cite{qH_SchuldEPL2017,qH_2BKB6arXiv2018,qH_BeheraQIP2017,qH_3BKB6arXiv2018,qH_BKB6arXiv2017,qH_Solano2arXiv2017} to name a few. 

Quantum chemistry has witnessed an upthrust in the application of quantum computers. Solving molecular problems using quantum mechanical laws makes it difficult because the interactions between large number of particles such as electrons cannot be handled by even powerful computers \cite{qH_Dirac1929}. Recent demonstrations of molecular simulations \cite{qH_KandalaNat2017} have paved the way for the study of complex molecules. Simulations are limited to small molecules due to hardware limitations. Various algorithms have been developed to calculate ground and excited states and mitigation of errors. In this paper, we demonstrate the use of one such algorithm known as Variational Quantum Eigensolver (VQE) for the simulation of $H^{-}$ ion. Initially developed by Peruzzo and McClean \cite{qH_PeruzzoNatComm2014}, VQE is a hybrid quantum-classical algorithm that uses both quantum and classical resources to solve the eigenvalue problem.

\begin{figure}[]
\includegraphics[width=1\linewidth]{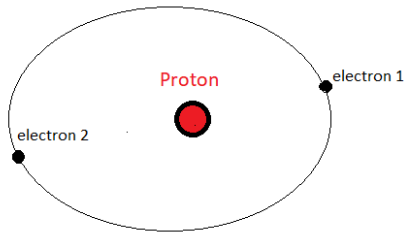}
\caption{\textbf{Structure of the Negative Hydrogen Ion }. Electron 1 is closer to the nucleus}
\label{qH_Fig1}
\end{figure}

The negative hydrogen ion is a three body system of a proton and two electrons where one electron is weakly attached to the nucleus. The structure of $H^{-}$ ion is shown in Fig. \ref{qH_Fig1}. One of the electrons is closer to the nucleus due to repulsion from another electron. Thus, the "inner" electron feels more charge than unity and the "outer" electron feels less charge than unity. None of the electrons can remain at the same radial distance from the nucleus due to Coloumb repulsion. It has an early history of theoretical research. It attracted attention as an application of quantum mechanics to a two electron system during the early development of quantum mechanics. It has an astrophysical importance as it is found in stars because of the presence of hydrogen and low energy electrons that can form a bound structure when an electron comes in the vicinity of the hydrogen atom \cite{qH_RauJAA1996}. The main source of opacity in the atmosphere of the Sun at red and infrared wavelengths was predicted due to the absorption by $H^{-}$ ion \cite{qH_WildtAJ1939}. At that time it was surprising that a system such as $H^{-}$, earlier believed to be unstable before the discovery of it's bound state could be a part of the solar spectrum, essential to sustain life on earth. It was discovered by G. Patrick Flanagan in 1983 to be present in the living fluids of all living organisms. There is very little dissociated hydrogen on earth and atmosphere. It's spectrum corresponds to infrared and visible wavelengths. Attempts were made to calculate the ground state energy of $H^{-}$ using perturbation and variational methods but these methods failed. The energy was calculated to be -0.375 Hartree which is greater than the ground state energy of hydrogen atom (-0.5 Hartree). To prove that it is a bound state the energy must be less than that of hydrogen atom. The dynamics of the system is such that it breaks up into proton + electron + electron at infinity when it reaches 2-3 eV above the threshold energy \cite{qH_RauJAA1996}. Thus, it becomes important to consider the electron-electron correlations. The wavefunction should describe the correlations efficiently in order to get a good measure of ground state energy.

$H^{-}$ was proved as a bound entity by Bethe in 1929 using Hylleraas wavefunction \cite{qH_BetheZphys1929}. The ground state energy was further improved by Chandrasekhar \cite{qH_ChandrasekharAPJ1944} by introducing a wavefunction of the following form,  

\begin{equation}
\psi(r1,r2)= e^{-ar_1-br_2}+e^{-br_1-ar_2}
\label{qH_Eq1}
\end{equation}

where a and b are variational parameters, representing the effective nuclear charges of the electrons. This wavefunction considers electron-electron correlation implicitly. The energy was found to be -0.51330 Hartree. A much better value was calculated by Chandrasekhar when he introduced electron-electron correlation explicitly in the improved wavefunction,

\begin{equation}
\psi =  \psi(r_{1},r_{2})(1+cr_{12})
\label{qH_Eq2}
\end{equation}

where c is the new variational parameter. The energy was found to be -0.52592 Hartree. Solving the Schrodinger using trial wavefunctions is a tedious task and becomes very complicated for multi-body systems because of the difficulty to guess the wavefunction that describes the system exactly. This difficulty can be addressed by quantum computers, which have been used for the simulation of various physical systems. The largest molecule to be simulated is $H_2O$ \cite{qH_TengarXiv2019}. This was done by using variational quantum eigensolver that is used to find eigenvalues of a matrix \cite{qH_PeruzzoNatComm2014,qH_GuzikNJP2016,qH_GuzikarXiv2019}. VQE is better than other quantum algorithms such as phase estimation due to it's high fidelity, robustness to errors and less resource requirement. It is based on the variational principle,

\begin{equation}
\frac{|\langle \psi| H | \psi \rangle |}{|\langle \psi | \psi \rangle|} \geq E
\label{qH_Eq3}
\end{equation}

where E is the minimum energy eigenvalue. The variational principle ensures that this expectation value is always greater than the smallest eigenvalue of H. Here, we report the calculation of the ground state energy of $H^{-}$ ion using VQE.
\section{Results}
\label{qed_section2}
The graphical representations of the results using various optimization methods obtained using QISKit are shown.

\begin{figure}
\centering
$\includegraphics[width=0.5\textwidth]{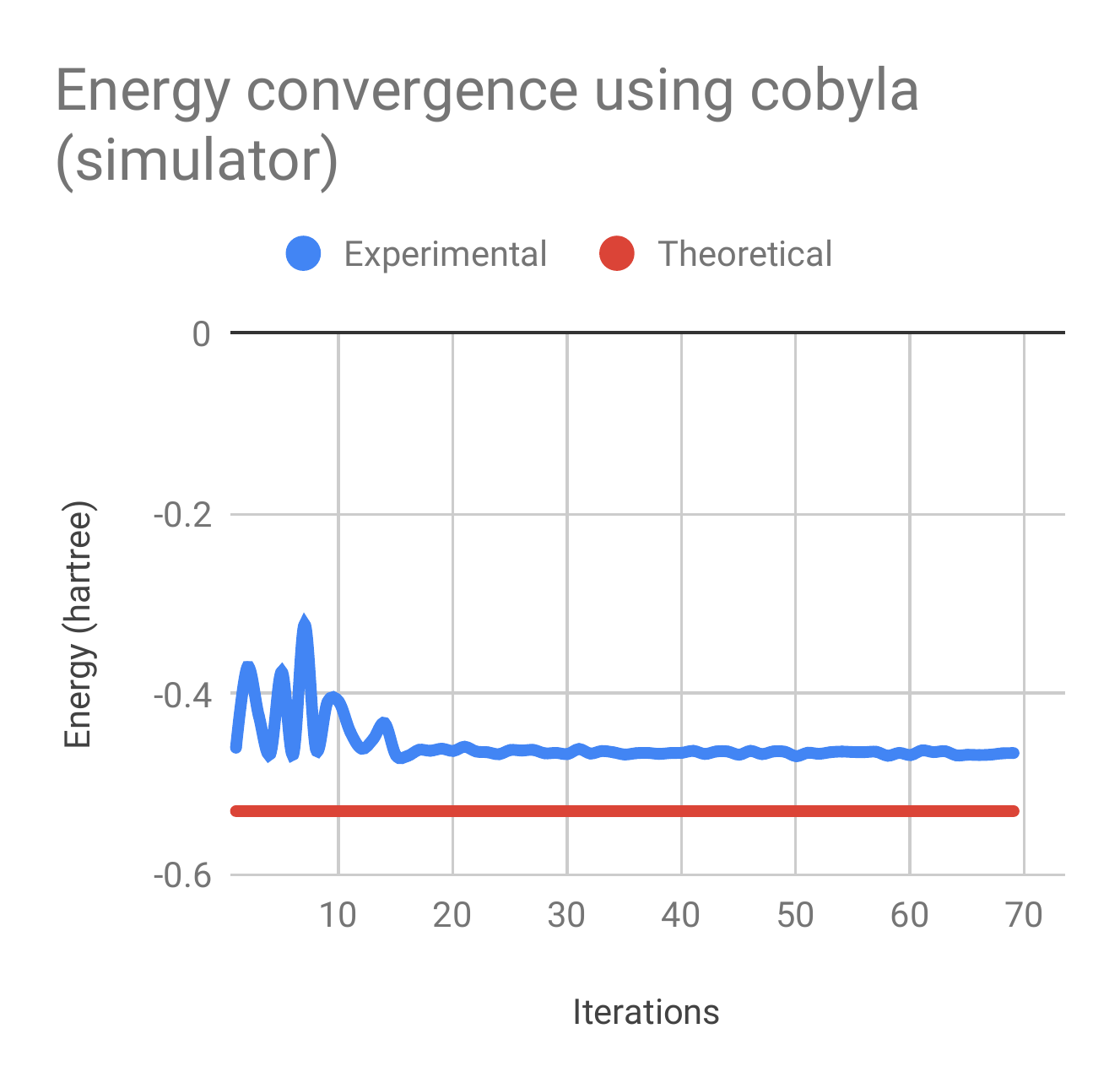}$
{\textbf{(a)}
$\includegraphics[width=0.5\textwidth]{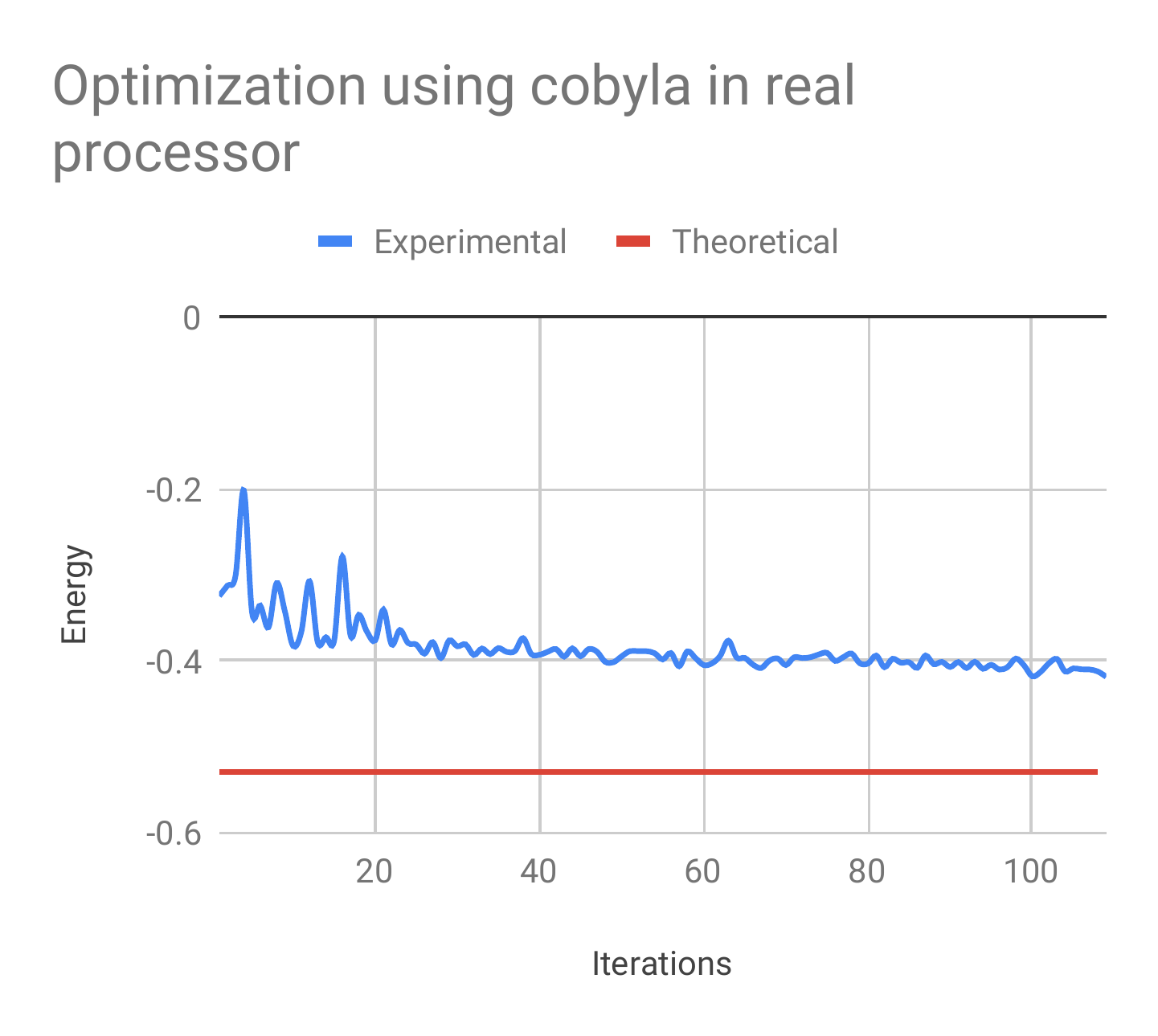}$
{\textbf{(b)}
\caption{\textbf{Energy optimization using Cobyla in QISKit aqua using ibmqx2 processor}. Gate depth is 3. The red line is the theoretical value of energy calculated from Chandrasekhar wavefunction which is -0.52952. Figures (a) and (b) show the results obtained from simulator and real device respectively.}}}
\label{qH_Fig2}
\end{figure}

\begin{figure}
\centering
$\includegraphics[width=0.5\textwidth]{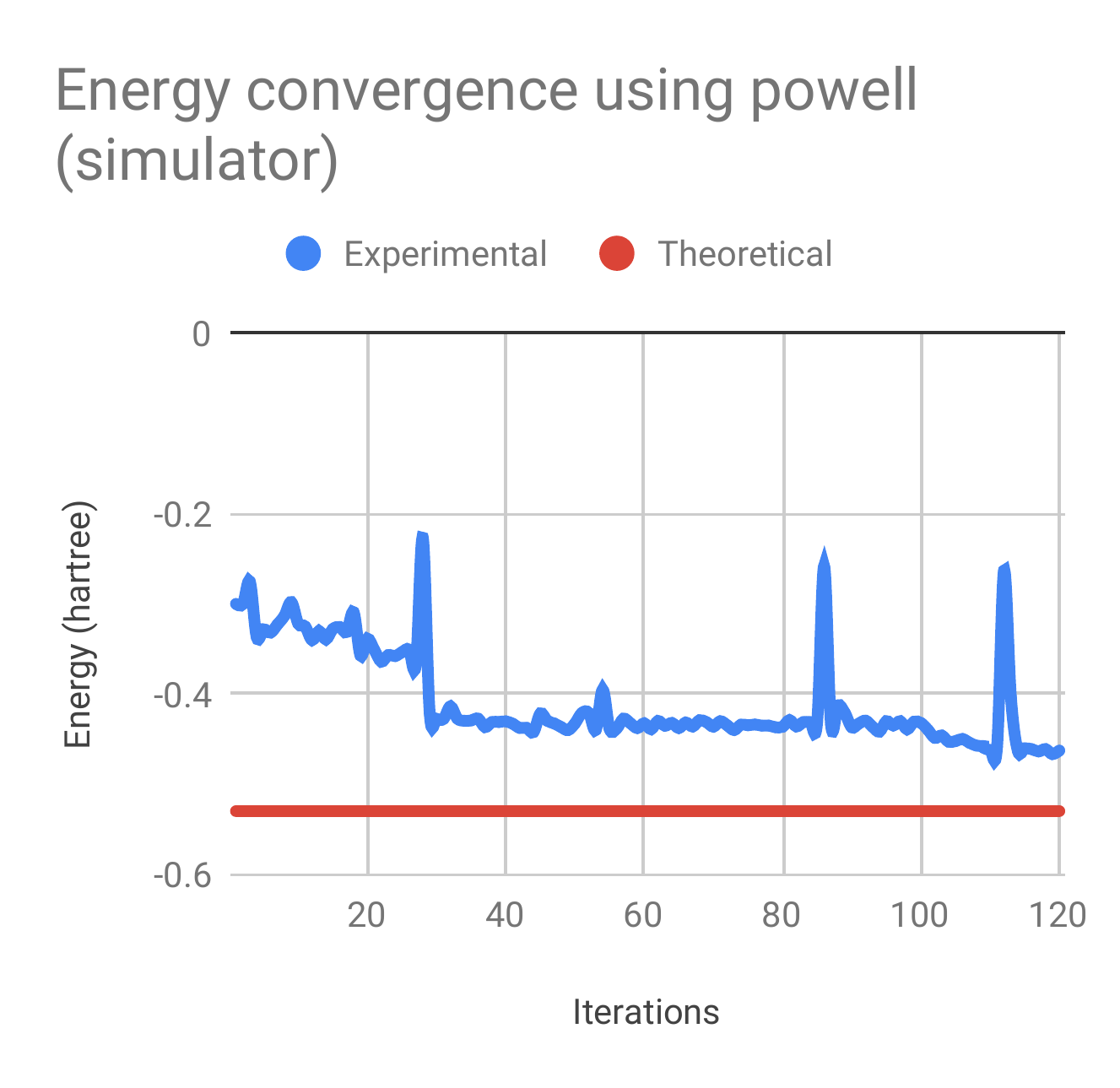}$
{\textbf{(c)}}
\label{qH_Fig3}
\includegraphics[width=0.5\textwidth]{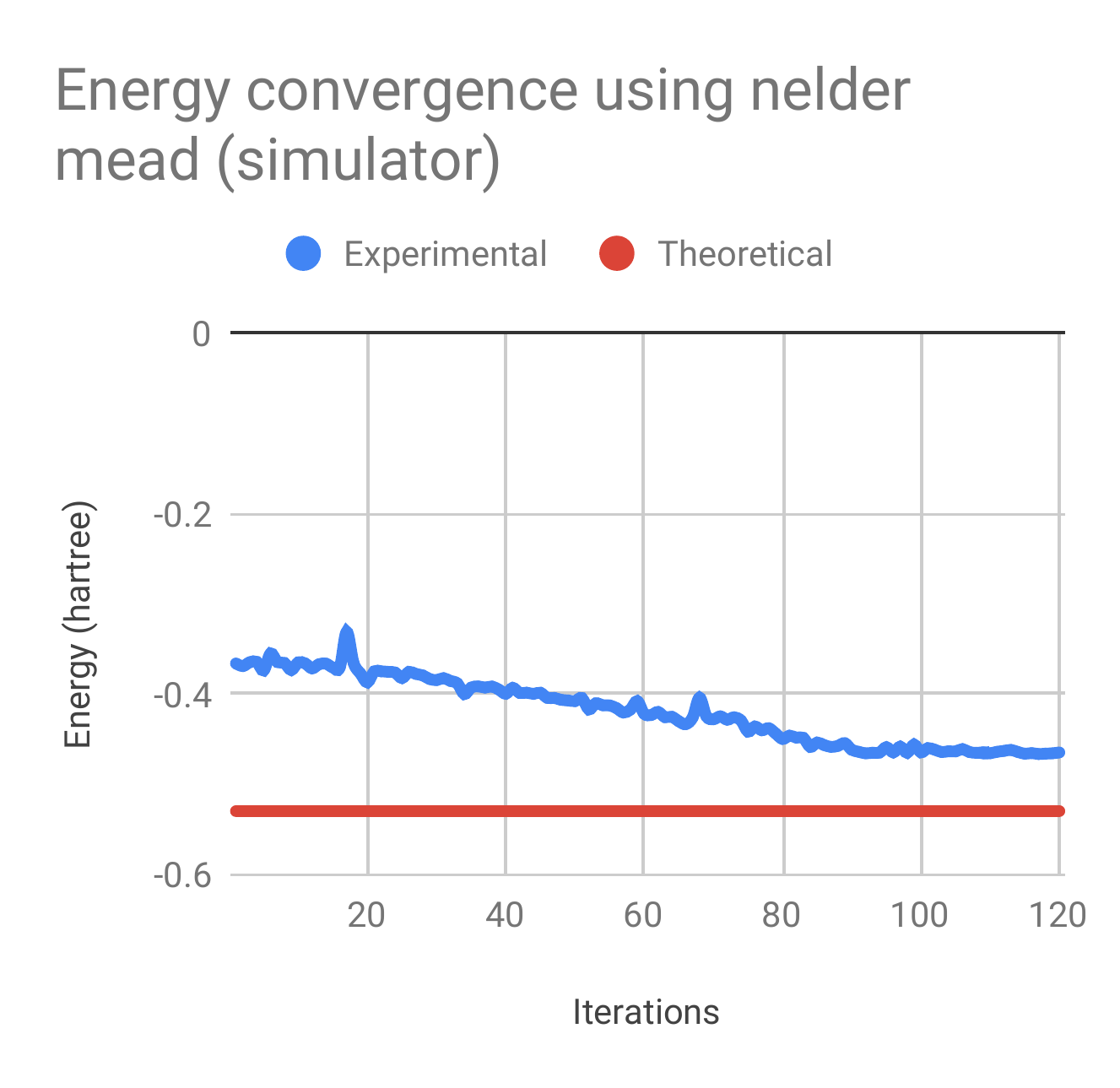}
{\textbf{(d)}
\caption{\textbf{Simulation results of energy optimization using Powell (c) and Nelder Mead (d) in QISKit aqua.} Gate depth is 3. The red line is the theoretical value of energy calculated from Chandrasekhar wavefunction which is 0.52952.}}
\label{qH_Fig2}
\end{figure}

The results are better than that calculated from the Hartree fock method (-0.375 Hartree). They converge to -0.468070601028 for Cobyla (simulator), -0.407087502741 for Cobyla (real processor, ibmqx2), -0.46513997401 using Powell (simulator) and -0.467324316239 for Nelder-Mead (simulator).
The results do not confirm the stability of $H^{-}$ as they are still greater than the ground state energy of hydrogen atom (-0.5 Hartree).
We run the VQE circuit in QISKit and calculate the expectation value of $Z_0I$ term. We use the same parameters of state preparation at which $Z_0I$ converged to calculate the expectation value of $IZ_1$ and $Z_0Z_1$. The probability of finding the electron in the first state is maximum. Thus, the lowest eigenvalue should be found using the set of parameters for which $Z_0I$ converges. Using Bravyi-Kitaev and Jordan-Wigner encoding the energy converged to -0.499711186 and -0.5339355468 (simulator) respectively much close to the theoretical value with an error of 0.8376\%. The energy converged to a lower value than the theoretical value but it is within the error bound. The details of the QISKit codes with results and experimental data are provided in the Supplementary information.

Run results obtained from ibmqx2 and ibmqx4 using different sets of parameters are given below.

\begin{figure}[H]
    \includegraphics[width=0.5\textwidth]{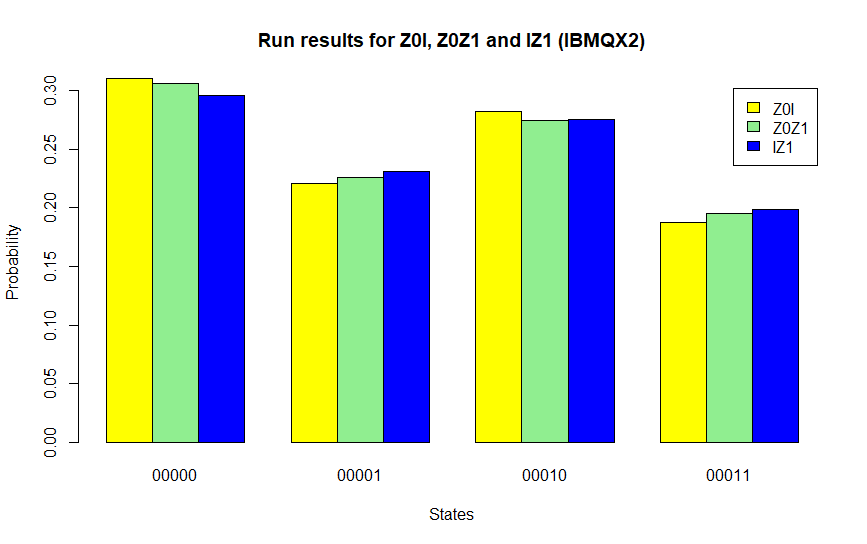}
\caption{Initial rotations of the parameters of 1st and 2nd Qubit are set to $\pi/2$ and final rotations are set to 0. Gate depth is 1. Number of shots = 8192}
\label{qH_Fig4}
\end{figure}

\begin{figure}[H]
        \includegraphics[width=0.5\textwidth]{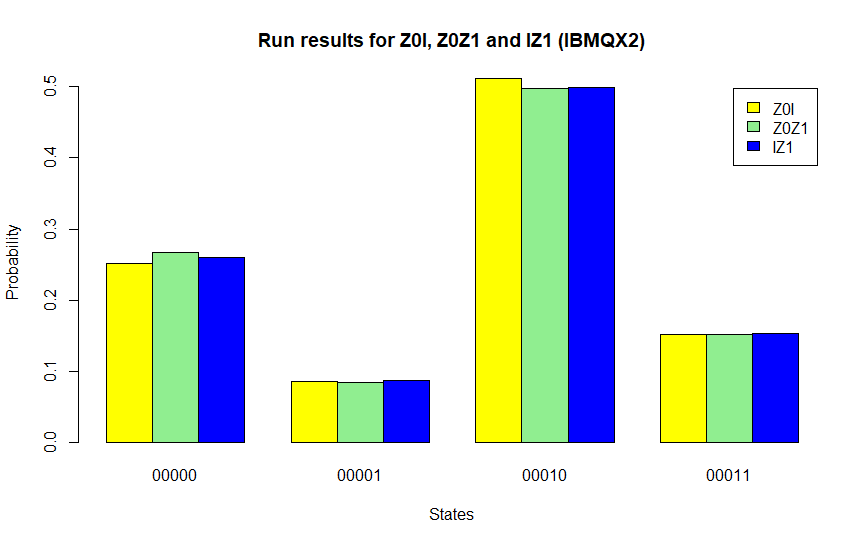}
\caption{Initial and final rotations of the parameters in 1st and 2nd Qubit are set to $\pi$. Gate depth is 1. Number of shots = 8192}
\label{qH_Fig5}
\end{figure}

\begin{figure}[H]
        \includegraphics[width=0.5\textwidth]{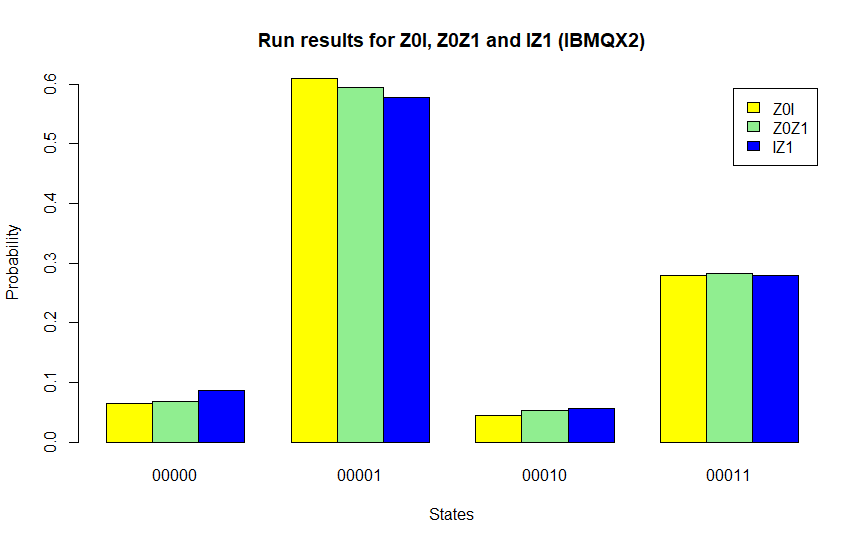}
\caption{Initial rotations of the parameters of 1st and 2nd Qubit are set to $\pi$ and final rotations are set to 0. Gate depth is 1. Number of shots = 8192}
\label{qH_Fig6}
\end{figure}

\begin{figure}
\includegraphics[width=0.5\textwidth]{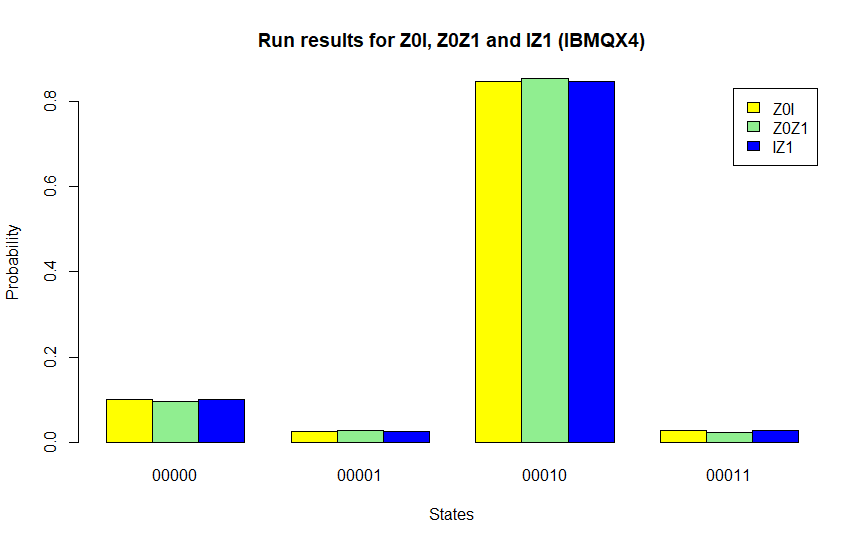}
{\caption{Parameters of both 1st and 2nd Qubit are set to $\pi$. Gate depth is 1. Number of shots = 8192}}
\label{qH_Fig7}
\end{figure}

The run results (Fig.\ref{qH_Fig4}) are obtained using ibmqx2. Initial rotations were set to $\pi/2$ and final rotations to 0 and we get the energy value -0.381156 Hartree. Setting initial and final rotations in both the qubits to $\pi$ (Fig.\ref{qH_Fig5}), we get the energy value -0.396531 Hartree. Setting initial rotations to $\pi$ and final rotations to 0 in both qubits (Fig.\ref{qH_Fig6}, gives the energy value -0.507891 Hartree.
Using ibmqx4, when the initial and final parameters in 1st and 2nd qubit are set to $\pi$ and 0 respectively we get the energy value -0.450297.

The energy value -0.507891 Hartree is lower than that of Hydrogen atom and thus confirms the bound state of $H^-$ ion.

\subsection{Variance}
We calculate the variance for the run result -0.507891 Hartree(ibmqx2)
\begin{equation}
    Var(H)= {|\langle\psi| H^2| \psi \rangle |} - {|\langle \psi| H | \psi \rangle |}^2
    \label{qH_Eq.4}
\end{equation}
\begin{itemize}
\item We first measure the variance of individual Pauli operators.
\end{itemize}
\begin{itemize}
\item The variance of the individual Pauli operators are then plucked in the above equation to obtain the variance.
\end{itemize}
The variance of individual Pauli operators is
\begin{equation}
|\langle\psi|\Delta P_i^2|\psi\rangle|  =\sum_i ({|\langle\psi|P_i^2|\psi\rangle |} - {|\langle\psi|P_i|\psi\rangle|}^2)
\end{equation}
where  $P_i$'s are individual Pauli operators.
Eq.4 becomes
\begin{equation}
Var(H)= \sum_{ij} h_{ij}^2(\Delta P_i^2) = 0.0870538
\end{equation}
\subsection{Conclusion}

Optimization of the VQE circuit using SciPy (Python module) produced good run results but insufficient to account for the bound state of $H^-$. This can be attributed to the possibility of energy convergance at a local minima. Run result of the VQE cirquit in the IBM Q Experience gave the energy value -0.507891 Hartree with an error of 3.3 \% from the theoretical value obtained using Chandrasekhar wavefunction. The entangled state of two electron system (correlated wavefunction of $H^-$) is efficiently realized in the IBM Quantum Computer.

\section*{Materials and Method}

\section{The Second-Quantized Hamiltonian} \label{qH_section3}
The second quantized Hamiltonian for fermions is given by,

\begin{equation}
H= \sum_{i,j}h_{ij} a_i^{\dagger}a_j +\frac{1}{2}\sum_{i,j,k,l}h_{ijkl} a_i^{\dagger}a_j^{\dagger}a_k a_l
\label{qH_Eq4}
\end{equation}
where $h_{ij}$ and $h_{ijkl}$ are one and two electron integrals given by
\begin{equation}
h_{ij} = \int d\Vec{r_1} \chi_i^*(\Vec{r_1}) (\frac{- \nabla_1^{2}}{2} -{\sum_ \sigma\frac{Z}{|\Vec{r_1}\ -R_\sigma|}})\chi_j(\Vec{r_1})
\label{qH_Eq5}
\end{equation} 
and
\begin{equation}
h_{ijkl} = \int \frac{ d\Vec{r_1} d\Vec{r_2} \chi_i^*(\Vec{r_1}) \chi_j^*(\Vec{r_2})}{|(r_1-r_2)|} \frac {{\chi_{k}(\Vec{r_2})}{\chi_l(\Vec{r_1})}}{2}
\label{qH_Eq6}
\end{equation} 

where ${\chi_i(\Vec{r_1})}$ is the $i$th spin orbital, Z is the nuclear charge, ${\Vec{r_i}}$ is the position of the $i$th electron, $r_{12}$ is the distance between the two points $r_{1}$ and $r_{2}$. $R_{\sigma}$ is the position of the nucleus. $a_{i}^{\dagger}$ and $a_{j}$ are fermionic creation and annihillation operators that follow the anti-commutation relations,

\begin{equation}
\{a_i^{\dagger},a_j\} = \delta_{ij}
\label{qH_Eq7}
\end{equation}
\begin{equation}
\{a_i,a_j\} = 0
\label{qH_Eq8}
\end{equation}

The fermionic creation operator increases the occupational number of an orbital by one and the annihilation operator decreases it by one. Using Jordan-Wigner or Bravyi-Kitaev transform, the fermionic Hamiltonian can be mapped to spin type Hamiltonian.

\label{qed_section4}
\section{The Jordan-Wigner transform}
\label{qH_section3}
Jordan-Wigner transform is a second-quantized encoding method to encode fermions into qubits. The mapping \cite{qH_bravyi2012} is given by,  
\begin{equation}
a{_i} = Q{_i}{\otimes}Z{_{i-1}}{\otimes}Z{_{i-2}}....{\otimes}Z{_0}
\label{qH_Eq9}
\end{equation}
\begin{equation}
a_{i}^{\dagger} = Q^{\dagger}{\otimes}Z{_{i-1}}{\otimes}Z{_{i-2}}....{\otimes}Z{_0}
\label{qH_Eq10}
\end{equation} where, 
\begin{equation}
Q =  \frac{X + iY}{2}
\label{qH_Eq11}
\end{equation}
\begin{equation}
Q{^\dagger} = \frac{X - iY}{2} 
\label{qH_Eq12}
\end{equation}

Each qubit stores the occupation number of the orbital.

\section{The Parity Transformation}

In the parity basis the qubit stores the parity of all occupied orbitals \cite{qH_bravyi2012}. The mapping is given by, $p_i$ = $\sum_{j}[\pi_n]_{ij}f_j$. This changes the occupation number basis state to it's corresponding parity basis state. The matrix form of $[\pi_n]_{ij}$ is given by,
\[
 \begin{split}
  \left[ {\begin{array}{cccc}
   1 & 0 & 0 & 0 \\
   1 & 1 & 0 & 0
   \\
   1 & 1 & 1 & 0
   \\
   1 & 1 & 1 & 1\\
  \end{array} } \right]
  \end{split}
 \] 
 
\section{The Bravyi Kitaev Transform}
The Bravyi Kitaev Transform is a midway between the Jordan Wigner and parity encoding. The orbitals store partial sums of occupation numbers \cite{qH_GuzikarXiv2019}. The qubit stores the parity of the set of occupation numbers corresponding to that set of orbitals. The qubits store occupation numbers when indices are even and parity when indices are odd. The transformation \cite{qH_bravyi2012} is given by, $b_i$ = $\sum_{j}[\beta_n]_{ij}f_j$, the matrix $[\beta_n]_{ij}$ for a 4 qubit system is,

\[
 \begin{split}
  \left[ {\begin{array}{cccc}
   1 & 0 & 0 & 0 \\
   1 & 1 & 0 & 0
   \\
   0 & 0 & 1 & 0
   \\
   1 & 1 & 1 & 1\\
  \end{array} } \right]
  \end{split}
 \] 
The relation of creation and annihilation operators \cite{qH_bravyi2012} is given by, 
\begin{equation}
a_{i}^{\dagger} = X_{U(i)}Q_i^{\dagger}{\otimes}Z_{p(i)}
\label{qH_Eq13}
\end{equation}

\begin{equation}
a_{i} = X_{U(i)}Q_i^{\dagger}{\otimes}Z_{p(i)}
\label{qH_Eq14}
\end{equation}

\section{Hamiltonian of the $H^{-}$ ion}\label{qH_section3}

The $H^{-}$ ion consists of two electrons, one of which is closer to the nucleus (Fig. \ref{qH_Fig1}). Thus, it has two states with one electron each. Thus the Hamiltonian given in Eq. \eqref{qH_Eq4} can be expanded using the Jordan-Wigner transform as \cite{qH_GuzikarXiv2019}, 

\begin{eqnarray}
H &= h_{00}(a_0^{\dagger}a_0) + h_{11}(a_1^{\dagger}a_1) + \frac{1}{2}h_{0101}(a_0^{\dagger}a_1^{\dagger}a_0a_1) \nonumber \\ &+ \frac{1}{2}h_{0110}(a_0^{\dagger}a_1^{\dagger}a_1a_0) + \frac{1}{2}h_{1001}(a_1^{\dagger}a_0^{\dagger}a_0a_1) \nonumber \\ & +\frac{1}{2}h_{1010}(a_1^{\dagger}a_0^{\dagger}a_1a_0)
\end{eqnarray}

Using the anti-commutation relations \eqref{qH_Eq7} and \eqref{qH_Eq8}, the Hamiltonian becomes,

\begin{eqnarray}
H &= h_{00}(a_{0}^{\dagger}a_{0}) + h_{11}(a_{1}^{\dagger}a_{1}) + h_{0101}(a_{0}^{\dagger}a_{1}^{\dagger}a_{0}a_{1}) \nonumber \\ &+ h_{0110}(a_{1}^{\dagger}a_{0}^{\dagger}a_{0}a_{1})
\end{eqnarray}

where we have used the fact that, $h_{0101} = h_{1010}$ and $h_{0110} = h_{1010}$. Using J-W transform \eqref{qH_Eq10}, we have,

\begin{equation}
a_0^{\dagger}a_0 = Q_0^{\dagger}Q_0, \nonumber
\end{equation}
\begin{equation}
a_1^{\dagger}a_1 = (Q_1^{\dagger}{\otimes}Z_0)(Q_1{\otimes}Z_0), \nonumber
\end{equation}
\begin{equation}
a_0^{\dagger}a_1^{\dagger}a_0a_1=(Q_0^{\dagger})(Q_1^{\dagger}{\otimes}Z_0)(Q_0)(Q_1{\otimes}Z_0), \nonumber
\end{equation}
\begin{equation}
a_1^{\dagger}a_0^{\dagger}a_0a_1 = (Q_1^{\dagger}{\otimes}Z_0)(Z_0)(Z_0)(Q_1{\otimes}Z_0)
\end{equation}

Making use of the tensor product relation $(A{\otimes}B)(C{\otimes}D) = (AC){\otimes}(BD)$, The Hamiltonian simplifies to,

\begin{eqnarray}
H &= \frac{1}{2}h_{00}{(1-Z_0)} + \frac{1}{2}h_{11}{(1-Z_1)} \nonumber \\&+\frac{1}{8}h_{0110}{(1-Z_0-Z_1+{Z_0}{\cdot Z_1)}}
\label{qH_Eq16}
\end{eqnarray}

Similarly using Bravyi-Kitaev transform the Hamiltonian for $H^{-}$ is

\begin{eqnarray}
H &= \frac{1}{2}h_{00}{(1-Z_0)} + \frac{1}{2}h_{11}{(1-Z_1Z_0)} \nonumber \\&+\frac{1}{8}h_{0110}{(1-Z_0+Z_1-{Z_0}{\cdot Z_1)}}
\end{eqnarray}
\label{qH_Eq16}

\section{Method of Simulation} \label{qH_Simulationmethod}
The method of simulation consists of 2 parts,

\textbf{Quantum}
\begin{itemize}
\item Prepare the state ${\psi}$, also known as the ansatz.
\item Measure the expectation value ${|\langle \psi| H | \psi \rangle |}$ using algorithms such as phase estimation or variational quantum eigensolver.
\end{itemize}

\textbf{Classical}
\begin{itemize}
\item Use a classical optimizer such as Nelder-Mead, Powell, Coybala and gradient descent methods for optimization.
\item Iterate until the energy converges.
\end{itemize}

\subsection{Variational Quantum Eigensolver}
\begin{figure}[]
\centering
\includegraphics[width=0.5\textwidth]{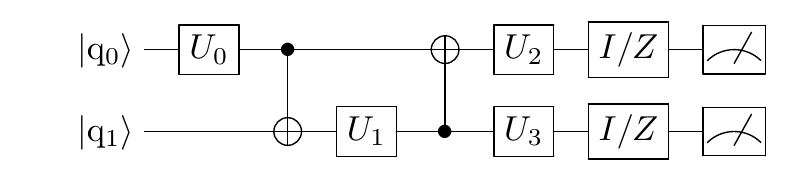}
\caption{\textbf{Quantum circuit illustrating the variational quantum eigensolver for two-qubit system.} The unitary operations $U_0$, $U_1$ and two $CNOT$ gates are used to create entanglement in the system for initial state preparation. $U_2$ and $U_3$ are then applied on the qubits $q_0$ and $q_1$ respectively for final rotations. Finally the terms of the Hamiltonian (Pauli matrices) are applied to calculate their expectation values.}
\label{qH_Figtwo}
\end{figure}

The quantum part of simulation is performed using variational quantum eigensolver as the algorithm to run the quantum subroutine. VQE was first demonstrated in 2014 \cite{qH_PeruzzoNatComm2014}. It has also been demonstrated in the Refs. \cite{qH_LeiWangarXiv2019,qH_MartinezarXiv2019,qH_CollessPRL2018}. The state preparation is done by entangling the qubits using various single qubit rotation gates to produce a complex state. VQE is capable of finding the ground state energies of small molecules using low depth circuits. It is based on the Rayleigh-Ritz variational principle,

\begin{equation}
{|\langle \psi ({\theta_i})| H | \psi ({\theta_i})\rangle |} \geq E_0.
\end{equation}

where $E_0$ is the ground state energy. The wave function is taken to be normalized. The circuit diagram of VQE \cite{qH_TengarXiv2019} for gate depth 1 is shown in Fig. \ref{qH_Figtwo}. The circuit consists of 2 parts, 

\textbf{Initial state preparation (ansatz) - }

\begin{itemize} 

\item The two qubit system is given initial rotations using the gates $U_0$ and $U_1$ followed by $CNOT$s. The unitary gates $U_{i}$ are of the form $R_z({\theta_i})R_x({\theta_i})R_z({\theta_i})$. The parameters ${\theta_i}$ have to be adjusted. This completes one layer of entanglement. The layers can be increased to produce more complex state as required. For an n-qubit system, the number of parameters for initial rotations would be 3n(n-1) and the number of unitary gates required is given n(n-1) \cite{qH_TengarXiv2019}. For a two-qubit system, we thus need six parameters and two unitary gates. It is then followed by final rotations $U_2$ and $U_3$.
\end{itemize}

\textbf{Measurement}
\begin{itemize}
\item The expectation values of each term in the Hamiltonian is measured one by one by introducing each Pauli term in the circuit after the state preparation as shown in Fig. \ref{qH_Figtwo}. 
\end{itemize}

\subsection{Optimization for Energy Convergence}
Optimization (classical part) is the last step for finding the minimum (ground state) energy. There are various optimization methods classified into two categories \cite{qH_GuzikarXiv2019},

\textbf{Direct Search Method}

Direct search algorithms do not make use of the gradient of the objective function e.g., particle swarm optimization, Nelder-Mead, Powell and Cobyla. These methods have been proven to be much better than gradient-based method. Nelder Mead is a good method for optimization. It is used in the fields of chemistry, medicine, science and technology. The method is derivative free and is used in systems where the functions are noisy and discontinuous such as parameter estimation and statistical problems. Powell method requires repeated line search minimization, which may be carried out using univariate gradient free, or gradient-based procedures. Cobyla constructs successive linear approximations of the objective function and constraints via a simplex of n+1 points (in n dimensions), and optimizes these approximations in a trust region at each step. 

\textbf{Gradient-based Method}
Gradient-based methods use the gradient of the objective function. Examples are simultaneous perturbation stochastic approximation (SPSA) algorithm, and L-BFGS-B. SPSA calculates the gradient by,

\begin{eqnarray}
G(\theta)&=\frac{1}{2} ({\langle \psi({\theta_i+{\pi/2}})| H | \psi ({\theta_i+{\pi/2}}\rangle})\nonumber \\& - {|\langle \psi ({\theta_i-{\pi/2}})| H| \psi ({\theta_i-{\pi/2}}\rangle})
\end{eqnarray}

where $G({\theta}$ is the gradient. The parameters are then updated according to whether the gradient decreases or increases. L-BFGS-B minimizes a differentiable scalar function f(x) over unconstrained values of the real-vector ``x'' \cite{qH_Malou2002}. Gradient-based optimization converges the function to a local minima, thus giving poor results. They are not used for functions with large number of parameters. Direct methods are much efficient and provide good values. But they are dependent on the system. Results vary with the number of qubits and gate depth. Here, we use Nelder-Mead, Cobyla and Powell methods and compare the results to depict which method is best suited for the calculation of ground state energy of $H^{-}$.

\subsection{Implementation}
In order to determine the ground state energy, we need the ``h'' values $h_{00}$, $h_{11}$ and $h_{0101}$ given in Eq. \eqref{qH_Eq16}. The ``h'' values are calculated by solving the one electron integrals (Eqs. \eqref{qH_Eq5} and \eqref{qH_Eq6}). They have been calculated in lecture 18 of the lecture series titled ``Quantum Principles'' \cite{qH_Lecture182014}. They are given as, $h_{00}$ = $h_{11}$ = 0.5, $h_{0101}$ = 0.625, $h_{00}$ and $h_{11}$ are equal as electrons are fermions (identical and indistinguishable particles). The expectation value of Hamiltonian is therefore,

\begin{eqnarray}
\langle \psi| H | \psi \rangle  &= \frac{1}{4}\langle \psi{(1-Z_0)}\psi \rangle + \frac{1}{4}\langle \psi{(1-Z_1)}\psi \rangle \nonumber \\ &+ \frac{5}{64}\langle \psi{(1-Z_0-Z_1-{Z_0}{\cdot Z_1)}\psi \rangle}
\label{qH_Eq18} 
\end{eqnarray}

The expectation of $Z_0$I, I$Z_1$ and $Z_{0}{{Z_1}}$ can be found by placing Z gates on qubits $q_0$ and $q_1$ after final rotation gates and measuring both the qubits.

In matrix form,
$Z_0$I = \[
 \begin{split}
  \left[ {\begin{array}{cccc}
   1 & 0 & 0 & 0 \\
   0 & 1 & 0 & 0
   \\
   0 & 0 & -1 & 0
   \\
   0 & 0 & 0 & -1\\
  \end{array} } \right]
  \end{split}
 \] 
the eigenvalues are 
\begin{itemize}
\item +1 for $|00\rangle$ and $|01\rangle$
\item -1 for $|10\rangle$ and $|11\rangle$
\end{itemize}
Expectation value is, $\langle \psi |Z_0I|\psi \rangle$ =  $P_{00}$ + $P_{01}$ - $P_{10}$  - $P_{11}$, where the eigenvalues have been multiplied as coefficients with respective eigenvectors. Similarly, 

I$Z_1$ = \[
\begin{split}
\left[ {\begin{array}{cccc}
   1 & 0 & 0 & 0 \\
   0 & -1 & 0 & 0
   \\
   0 & 0 & 1 & 0
   \\
   0 & 0 & 0 & -1\\
  \end{array} } \right]
  \end{split}
 \] 
the eigenvalues are 

\begin{itemize}
\item +1 for $|00\rangle$ and $|10\rangle$
\item -1 for $|01\rangle$ and $|11\rangle$
\end{itemize}

$\langle \psi |IZ_1| \psi \rangle$ =  $P_{00}$ - $P_{01}$ + $P_{10}$ - $P_{11}$ and
 
$Z_0 Z_1$ = \[
 \begin{split}
  \left[ {\begin{array}{cccc}
   1 & 0 & 0 & 0 \\
   0 & -1 & 0 & 0
   \\
   0 & 0 & -1 & 0
   \\
   0 & 0 & 0 & 1\\
  \end{array} } \right]
  \end{split}
 \] 
the eigenvalues are 
\begin{itemize}
\item +1 for $|00\rangle$ and $|11\rangle$
\item -1 for $|01\rangle$ and $|10\rangle $
\end{itemize}
\begin{equation}
\langle \psi |Z_0Z_1|\psi \rangle = P_{00} - P_{01} - P_{10}  + P_{11}
\end{equation}

The initial parameters may be chosen at random. It can be adjusted according to the measured expectation value. We depict a comparison between various optimization methods. The optimization was performed by running codes in QISKit and running program in QISKit aqua.

\section*{Data availability}
Data are available to any reader upon reasonable request.

\section*{Acknowledgments}
\label{qlock_acknowledgments}
S.K. would like to thank Indian Institute of Science Education and Research Kolkata for providing hospitality during the course of the project. B.K.B. acknowledges the support of IISER-K Institute Fellowship. The authors acknowledge the support of IBM Quantum Experience for producing experimental results. The views expressed are those of the authors and do not reflect the official policy or position of IBM or the IBM Quantum Experience team. The authors thank Dheerendra Singh (IPhD Student at IISER Kolkata) for useful discussions.

\section*{Author contributions}
S.K. has done theoretical analysis and developed the protocol. S.K. and B.K.B. have analyzed and designed the quantum circuits. S.K., R.P.S. and B.K.B. have implemented the circuits on IBM quantum experience platform and performed the experiments. R.P.S. has written the circuit codes, run the optimization routine and quantum chemistry program for energy convergence in QISKit Terra and QISKit Aqua. S.K. and B.K.B. contributed to the composition of the manuscript. B.K.B has supervised the project. P.K.P. thoroughly checked and reviewed the manuscript. S.K., R.P.S. and B.K.B. have completed the project under the guidance of P.K.P.  
\section*{Competing interests}
The authors declare no competing financial as well as non-financial interests.

\clearpage
\section{Supplementary Information: QISKIT codes for optimization.}

For Energy estimation $Z_0I$, $IZ_1$ and $Z_0Z_1$ terms of Hamiltonian respectively are coded then optimized through scipy.optimize which contain 'powell', 'nelder-mead' or 'cobyla'. The QASM code for the same is as follows:\\ 
Code for Z1($Z_0I$):
\begin{lstlisting}[language=Python]
# -*- coding: utf-8 -*-
"""
Created on Wed Feb 13 20:04:16 2019

@author: Rahul
"""

from scipy.optimize import minimize
from qiskit import QuantumCircuit, ClassicalRegister, QuantumRegister
import numpy as np
from qiskit import execute
from qiskit import BasicAer
backend = BasicAer.get_backend('qasm_simulator')
T=8192
def Z1(theta):
    # Create a Quantum Register called "q" with 3 qubits
    q = QuantumRegister(2)
        
    # Create a Classical Register called "c" with 3 bits
    c = ClassicalRegister(2)
    qc = QuantumCircuit(q,c)
    
    qc.u1(theta[0],q[0])
    qc.u3(theta[1],-np.pi/2,np.pi/2,q[0])
    qc.u1(theta[2],q[0])
    qc.cx(q[0], q[1])
    qc.u1(theta[3],q[1])
    qc.u3(theta[4],-np.pi/2,np.pi/2,q[1])
    qc.u1(theta[5],q[1])
    qc.cx(q[1], q[0])
    qc.u1(theta[6],q[0])
    qc.u1(theta[7],q[1])
    qc.u3(theta[8],-np.pi/2,np.pi/2,q[0])
    qc.u3(theta[9],-np.pi/2,np.pi/2,q[1])
    qc.u1(theta[10],q[0])
    qc.u1(theta[11],q[1])
    qc.z(q[0])
    
    qc.measure(q[0], c[0])
    qc.measure(q[1], c[1])
    
    #print(qc)
    #print(i)
    shots = T      # Number of shots to run the program (experiment); maximum is 8192 shots.
    max_credits = 3        # Maximum number of credits to spend on executions. 

    job_hpc = execute(qc, backend=backend, shots=shots, max_credits=max_credits)
    result_hpc = job_hpc.result()
    counts11 = result_hpc.get_counts(qc) 
    #print(counts11)
    Z=0
    if '00' in list(counts11):
        Z=Z+counts11['00']/T
        #print(Z)
    if '01' in list(counts11):
        Z=Z+counts11['01']/T
        #print(Z)
    if '10' in list(counts11):
        Z=Z-counts11['10']/T
        #print(Z)
    if '11' in list(counts11):
        Z=Z-counts11['11']/T
        #print(Z)
        
    return Z
theta0=[0,np.pi/2,0,0,np.pi/2,0,0,0,np.pi/2,np.pi/2,0,0]
res = minimize(Z1, theta0, method='powell',options={'xtol': 1e-8 , 'disp': True})
print(res,'Z1')
#x=Z1([ 1.53637770e-03,  1.55027283e+00,  8.09473120e-04,  1.66340534e-04,-8.32420733e-03, -2.28166637e-05, -8.94140209e-04,  5.61023344e-04,1.77675920e+00,  1.58425510e+00,  1.09455026e-03,  8.51956374e-04])  
#print(x)
\end{lstlisting}
Code for Z2($IZ_1$):
\begin{lstlisting}[language=Python]
# -*- coding: utf-8 -*-
"""
Created on Wed Feb 13 20:04:17 2019

@author: Rahul
"""

from scipy.optimize import minimize
from qiskit import QuantumCircuit, ClassicalRegister, QuantumRegister
import numpy as np
from qiskit import execute
from qiskit import BasicAer
backend = BasicAer.get_backend('qasm_simulator')
T=8192
def Z2(theta):
    # Create a Quantum Register called "q" with 3 qubits
    q = QuantumRegister(2)
        
    # Create a Classical Register called "c" with 3 bits
    c = ClassicalRegister(2)
    qc = QuantumCircuit(q,c)
    
    qc.u1(theta[0],q[0])
    qc.u3(theta[1],-np.pi/2,np.pi/2,q[0])
    qc.u1(theta[2],q[0])
    qc.cx(q[0], q[1])
    qc.u1(theta[3],q[1])
    qc.u3(theta[4],-np.pi/2,np.pi/2,q[1])
    qc.u1(theta[5],q[1])
    qc.cx(q[1], q[0])
    qc.u1(theta[6],q[0])
    qc.u1(theta[7],q[1])
    qc.u3(theta[8],-np.pi/2,np.pi/2,q[0])
    qc.u3(theta[9],-np.pi/2,np.pi/2,q[1])
    qc.u1(theta[10],q[0])
    qc.u1(theta[11],q[1])
    qc.z(q[1])
    
    qc.measure(q[0], c[0])  
    qc.measure(q[1], c[1])
    
    #print(qc)
    #print(i)
    shots = T           # Number of shots to run the program (experiment); maximum is 8192 shots.
    max_credits = 3        # Maximum number of credits to spend on executions. 

    job_hpc = execute(qc, backend=backend, shots=shots, max_credits=max_credits)
    result_hpc = job_hpc.result()
    counts22 = result_hpc.get_counts(qc)
    
    Z=0
    if '00' in list(counts22):
        Z=Z+counts22['00']/T
        print(Z)
    if '01' in list(counts22):
        Z=Z-counts22['01']/T
        print(Z)
    if '10' in list(counts22):
        Z=Z+counts22['10']/T
        print(Z)
    if '11' in list(counts22):
        Z=Z-counts22['11']/T
        print(Z)
        
    return Z
theta0=[0,np.pi/2,0,0,np.pi/2,0,0,0,np.pi/2,np.pi/2,0,0]
res = minimize(Z2, theta0, method='nelder-mead',options={'xtol': 1e-8 , 'disp': True})
print(res,'Z2')
#z=Z2([-5.64200351e-01,  1.61554986e+00,  1.56821823e+00,  1.61219634e-03, 1.81902336e+00,  5.96777406e+00, -6.66574506e-03,  4.63725496e+00, 1.32156756e+00,  3.78827977e-01,  9.59049221e+00,  3.90466495e+00])
#print(z)
\end{lstlisting}
Code for Z3($Z_0Z_I$):
\begin{lstlisting}[language=Python]
# -*- coding: utf-8 -*-
"""
Created on Wed Feb 13 20:04:15 2019

@author: Rahul
"""

from scipy.optimize import minimize
from qiskit import QuantumCircuit, ClassicalRegister, QuantumRegister
import numpy as np
from qiskit import execute
from qiskit import IBMQ
backend =IBMQ.get_backend('ibmqx4')
T=8192
def Z3(theta):
    # Create a Quantum Register called "q" with 3 qubits
    q = QuantumRegister(2)
        
    # Create a Classical Register called "c" with 3 bits
    c = ClassicalRegister(2)
    qc = QuantumCircuit(q,c)
    
    qc.u1(theta[0],q[0])
    qc.u3(theta[1],-np.pi/2,np.pi/2,q[0])
    qc.u1(theta[2],q[0])
    qc.cx(q[0], q[1])
    qc.u1(theta[3],q[1])
    qc.u3(theta[4],-np.pi/2,np.pi/2,q[1])
    qc.u1(theta[5],q[1])
    qc.cx(q[1], q[0])
    qc.u1(theta[6],q[0])
    qc.u1(theta[7],q[1])
    qc.u3(theta[8],-np.pi/2,np.pi/2,q[0])
    qc.u3(theta[9],-np.pi/2,np.pi/2,q[1])
    qc.u1(theta[10],q[0])
    qc.u1(theta[11],q[1])
    qc.z(q[0])
    qc.z(q[1])
    
    qc.measure(q[0], c[0])
    qc.measure(q[1], c[1])  
    
    #print(qc)
    #print(i)
    shots= T          # Number of shots to run the program (experiment); maximum is 8192 shots.
    max_credits = 3        # Maximum number of credits to spend on executions. 

    job_hpc = execute(qc, backend=backend, shots=shots, max_credits=max_credits)
    result_hpc = job_hpc.result()
    counts12 = result_hpc.get_counts(qc)
    
    Z=0
    if '00' in list(counts12):
        Z=counts12['00']/T
        print(Z)
    if '01' in list(counts12):
        Z=Z-counts12['01']/T
        print(Z)
    if '10' in list(counts12):
        Z=Z-counts12['10']/T
        print(Z)
    if '11' in list(counts12):
        Z=Z+counts12['11']/T
        print(Z)
    return Z
theta0=[0,np.pi/2,0,0,np.pi/2,0,0,0,np.pi/2,np.pi/2,0,0]
res = minimize(Z3, theta0, method='nelder-mead',options={'xtol': 1e-8 , 'disp': True})
print(res,'Z3')
#y=Z3([-7.63609374, -0.31633082,  2.15909817,  4.31092763,  1.57470161, 1.6111563 , -0.13041641,  0.08880943,  1.58865914,  1.50475982, 0.59323763,  2.86816974])
#print(y)
\end{lstlisting}
The tabulated data of convergence is as shown below.
\begin{table}[H]
\centering
\caption{For Nelder Mead } \label{tab:title} 
\begin{tabular}{|l|l|l|}
\hline
Iterations & Theoretical & Experimental  \\ \hline
1          & -0.52952    & -0.3658118514 \\ \hline
2          & -0.52952    & -0.3687388385 \\ \hline
3          & -0.52952    & -0.364456499  \\ \hline
4          & -0.52952    & -0.3640980715 \\ \hline
5          & -0.52952    & -0.3735459878 \\ \hline
6          & -0.52952    & -0.3547868431 \\ \hline
7          & -0.52952    & -0.3642996161 \\ \hline
8          & -0.52952    & -0.3652836663 \\ \hline
9          & -0.52952    & -0.3730938032 \\ \hline
10         & -0.52952    & -0.3646086323 \\ \hline
11         & -0.52952    & -0.3657974241 \\ \hline
12         & -0.52952    & -0.3712501318 \\ \hline
13         & -0.52952    & -0.3667064972 \\ \hline
14         & -0.52952    & -0.3660282661 \\ \hline
15         & -0.52952    & -0.3702449038 \\ \hline
16         & -0.52952    & -0.3702913827 \\ \hline
17         & -0.52952    & -0.3308180563 \\ \hline
18         & -0.52952    & -0.3653252632 \\ \hline
19         & -0.52952    & -0.3770401351 \\ \hline
20         & -0.52952    & -0.3867578684 \\ \hline
21         & -0.52952    & -0.3746037324 \\ \hline
22         & -0.52952    & -0.3745729229 \\ \hline
23         & -0.52952    & -0.375111539  \\ \hline
24         & -0.52952    & -0.3759054584 \\ \hline
25         & -0.52952    & -0.3819807294 \\ \hline
26         & -0.52952    & -0.3751555337 \\ \hline
27         & -0.52952    & -0.3774191599 \\ \hline
28         & -0.52952    & -0.3790968918 \\ \hline
29         & -0.52952    & -0.3830969824 \\ \hline
30         & -0.52952    & -0.3842554535 \\ \hline
31         & -0.52952    & -0.3821377446 \\ \hline
32         & -0.52952    & -0.3852416368 \\ \hline
33         & -0.52952    & -0.387955228  \\ \hline
34         & -0.52952    & -0.3997053027 \\ \hline
35         & -0.52952    & -0.3925481822 \\ \hline
36         & -0.52952    & -0.3912610618 \\ \hline
37         & -0.52952    & -0.3925515116 \\ \hline
38         & -0.52952    & -0.39142132   \\ \hline
39         & -0.52952    & -0.3948359592 \\ \hline
40         & -0.52952    & -0.3993956196 \\ \hline
41         & -0.52952    & -0.3929185069 \\ \hline
42         & -0.52952    & -0.3984600897 \\ \hline
43         & -0.52952    & -0.3981955975 \\ \hline
44         & -0.52952    & -0.3995114402 \\ \hline
45         & -0.52952    & -0.3982939261 \\ \hline
46         & -0.52952    & -0.4039865324 \\ \hline
47         & -0.52952    & -0.4040763391 \\ \hline
48         & -0.52952    & -0.4061567579 \\ \hline
49         & -0.52952    & -0.4068460868 \\ \hline
50         & -0.52952    & -0.4075465596 \\ \hline
\end{tabular}
\end{table}

\begin{table}[H]
\begin{tabular}{|l|l|l|}
\hline
Iterations & Theoretical & Experimental  \\ \hline
51         & -0.52952    & -0.4042369078 \\ \hline
52         & -0.52952    & -0.4167492066 \\ \hline
53         & -0.52952    & -0.4100107529 \\ \hline
54         & -0.52952    & -0.4121749407 \\ \hline
55         & -0.52952    & -0.4123393275 \\ \hline
56         & -0.52952    & -0.4153778284 \\ \hline
57         & -0.52952    & -0.4201981629 \\ \hline
58         & -0.52952    & -0.416993412  \\ \hline
59         & -0.52952    & -0.4075684502 \\ \hline
60         & -0.52952    & -0.4216554432 \\ \hline
61         & -0.52952    & -0.4229979668 \\ \hline
62         & -0.52952    & -0.4196147123 \\ \hline
63         & -0.52952    & -0.4253503764 \\ \hline
64         & -0.52952    & -0.4249917301 \\ \hline
65         & -0.52952    & -0.4299162527 \\ \hline
66         & -0.52952    & -0.4330937475 \\ \hline
67         & -0.52952    & -0.4257880472 \\ \hline
68         & -0.52952    & -0.4038551852 \\ \hline
69         & -0.52952    & -0.424561212  \\ \hline
70         & -0.52952    & -0.4276609296 \\ \hline
71         & -0.52952    & -0.4244919108 \\ \hline
72         & -0.52952    & -0.4281787243 \\ \hline
73         & -0.52952    & -0.4255140557 \\ \hline
74         & -0.52952    & -0.4291707672 \\ \hline
75         & -0.52952    & -0.4413052833 \\ \hline
76         & -0.52952    & -0.4357576628 \\ \hline
77         & -0.52952    & -0.4398903991 \\ \hline
78         & -0.52952    & -0.4376459041 \\ \hline
79         & -0.52952    & -0.4434627664 \\ \hline
80         & -0.52952    & -0.4494898276 \\ \hline
81         & -0.52952    & -0.445895972  \\ \hline
82         & -0.52952    & -0.4480489295 \\ \hline
83         & -0.52952    & -0.448329137  \\ \hline
84         & -0.52952    & -0.4582959119 \\ \hline
85         & -0.52952    & -0.4535441657 \\ \hline
86         & -0.52952    & -0.4562919823 \\ \hline
87         & -0.52952    & -0.458321483  \\ \hline
88         & -0.52952    & -0.4572042524 \\ \hline
89         & -0.52952    & -0.4542379797 \\ \hline
90         & -0.52952    & -0.4615284535 \\ \hline
91         & -0.52952    & -0.4637654447 \\ \hline
92         & -0.52952    & -0.4654706567 \\ \hline
93         & -0.52952    & -0.4647630364 \\ \hline
94         & -0.52952    & -0.4646672379 \\ \hline
95         & -0.52952    & -0.4587078335 \\ \hline
96         & -0.52952    & -0.4644073631 \\ \hline
97         & -0.52952    & -0.458071464  \\ \hline
98         & -0.52952    & -0.4648104928 \\ \hline
99         & -0.52952    & -0.455790524  \\ \hline
100        & -0.52952    & -0.4643548465 \\ \hline
\end{tabular}
\end{table}

\begin{table}[H]
\begin{tabular}{|l|l|l|}
\hline
Iterations & Theoretical & Experimental  \\ \hline
101        & -0.52952    & -0.4597123945 \\ \hline
102        & -0.52952    & -0.4614880122 \\ \hline
103        & -0.52952    & -0.4640232781 \\ \hline
104        & -0.52952    & -0.4631510063 \\ \hline
105        & -0.52952    & -0.4633038065 \\ \hline
106        & -0.52952    & -0.4607956644 \\ \hline
107        & -0.52952    & -0.4640445423 \\ \hline
108        & -0.52952    & -0.4648928415 \\ \hline
109        & -0.52952    & -0.464328522  \\ \hline
109        & -0.52952    & -0.4654582708 \\ \hline
110        & -0.52952    & -0.4653556353 \\ \hline
111        & -0.52952    & -0.4637653123 \\ \hline
112        & -0.52952    & -0.4627551564 \\ \hline
113        & -0.52952    & -0.4616840943 \\ \hline
114        & -0.52952    & -0.4639483874 \\ \hline
115        & -0.52952    & -0.4658633096 \\ \hline
116        & -0.52952    & -0.4652171301 \\ \hline
117        & -0.52952    & -0.4662359862 \\ \hline
118        & -0.52952    & -0.4657969867 \\ \hline
119        & -0.52952    & -0.4654209807 \\ \hline
120        & -0.52952    & -0.4644860719 \\ \hline
121         & -0.52952    & -0.4655182454 \\ \hline
122        & -0.52952    & -0.4639068364 \\ \hline
123        & -0.52952    & -0.4650023139 \\ \hline
124        & -0.52952    & -0.4634304145 \\ \hline
125        & -0.52952    & -0.4670682595 \\ \hline
\end{tabular}
\end{table}

\begin{table}[H]
\centering
\caption{For Cobyla on real Chip } \label{tab:title} 
\begin{tabular}{|l|l|l|}
\hline
Iterations & Theoretical & Experimental  \\ \hline
1          & -0.52952    & -0.3242458015 \\ \hline
2          & -0.52952    & -0.312141158  \\ \hline
3          & -0.52952    & -0.3000890929 \\ \hline
4          & -0.52952    & -0.2012450742 \\ \hline
5          & -0.52952    & -0.3431368385 \\ \hline
6          & -0.52952    & -0.3354519405 \\ \hline
7          & -0.52952    & -0.3605707275 \\ \hline
8          & -0.52952    & -0.3092321975 \\ \hline
9          & -0.52952    & -0.3421126117 \\ \hline
10         & -0.52952    & -0.382317455  \\ \hline
11         & -0.52952    & -0.365973851  \\ \hline
12         & -0.52952    & -0.3070352964 \\ \hline
13         & -0.52952    & -0.378163368  \\ \hline
14         & -0.52952    & -0.372308508  \\ \hline
15         & -0.52952    & -0.3760271897 \\ \hline
16         & -0.52952    & -0.2783118205 \\ \hline
17         & -0.52952    & -0.3703690787 \\ \hline
18         & -0.52952    & -0.3461754311 \\ \hline
19         & -0.52952    & -0.3659191554 \\ \hline
20         & -0.52952    & -0.3757416573 \\ \hline
21         & -0.52952    & -0.3401756666 \\ \hline
22         & -0.52952    & -0.3805747747 \\ \hline
23         & -0.52952    & -0.3638238258 \\ \hline
24         & -0.52952    & -0.3789247876 \\ \hline
25         & -0.52952    & -0.3807448887 \\ \hline
26         & -0.52952    & -0.3913352043 \\ \hline
27         & -0.52952    & -0.3779779816 \\ \hline
28         & -0.52952    & -0.3963448298 \\ \hline
29         & -0.52952    & -0.3759803137 \\ \hline
30         & -0.52952    & -0.3827794955 \\ \hline
31         & -0.52952    & -0.3805431712 \\ \hline
32         & -0.52952    & -0.3928037149 \\ \hline
33         & -0.52952    & -0.385364926  \\ \hline
34         & -0.52952    & -0.3918327581 \\ \hline
35         & -0.52952    & -0.3848496156 \\ \hline
36         & -0.52952    & -0.3892537216 \\ \hline
37         & -0.52952    & -0.3883246723 \\ \hline
38         & -0.52952    & -0.3732819491 \\ \hline
39         & -0.52952    & -0.3917883664 \\ \hline
40         & -0.52952    & -0.3926158442 \\ \hline
41         & -0.52952    & -0.3891585442 \\ \hline
42         & -0.52952    & -0.3859512631 \\ \hline
43         & -0.52952    & -0.3947456284 \\ \hline
44         & -0.52952    & -0.3852036904 \\ \hline
45         & -0.52952    & -0.3942924663 \\ \hline
46         & -0.52952    & -0.3859246281 \\ \hline
47         & -0.52952    & -0.3890963958 \\ \hline
48         & -0.52952    & -0.4008480231 \\ \hline
49         & -0.52952    & -0.4013480612 \\ \hline
50         & -0.52952    & -0.3944512178 \\ \hline
\end{tabular}
\end{table}

\begin{table}[H]
\begin{tabular}{|l|l|l|}
\hline
Iterations & Theoretical & Experimental  \\ \hline
51         & -0.52952    & -0.3884351222 \\ \hline
52         & -0.52952    & -0.3884351222 \\ \hline
53         & -0.52952    & -0.3884351222 \\ \hline
54         & -0.52952    & -0.3907311105 \\ \hline
55         & -0.52952    & -0.3980456027 \\ \hline
56         & -0.52952    & -0.3906448114 \\ \hline
57         & -0.52952    & -0.4059641886 \\ \hline
58         & -0.52952    & -0.388795591  \\ \hline
59         & -0.52952    & -0.3965149438 \\ \hline
60         & -0.52952    & -0.4046238846 \\ \hline
61         & -0.52952    & -0.4025665527 \\ \hline
62         & -0.52952    & -0.3938826373 \\ \hline
63         & -0.52952    & -0.3762519993 \\ \hline
64         & -0.52952    & -0.395794006  \\ \hline
65         & -0.52952    & -0.3963359515 \\ \hline
66         & -0.52952    & -0.4041835107 \\ \hline
67         & -0.52952    & -0.4079657663 \\ \hline
68         & -0.52952    & -0.3994302984 \\ \hline
69         & -0.52952    & -0.3969173201 \\ \hline
70         & -0.52952    & -0.4045947654 \\ \hline
71         & -0.52952    & -0.3955795003 \\ \hline
72         & -0.52952    & -0.396464158  \\ \hline
73         & -0.52952    & -0.3954576879 \\ \hline
74         & -0.52952    & -0.3922859202 \\ \hline
75         & -0.52952    & -0.3903440066 \\ \hline
76         & -0.52952    & -0.3997641322 \\ \hline
77         & -0.52952    & -0.3953649947 \\ \hline
78         & -0.52952    & -0.3912133919 \\ \hline
79         & -0.52952    & -0.4025246453 \\ \hline
80         & -0.52952    & -0.4028140875 \\ \hline
81         & -0.52952    & -0.3936148616 \\ \hline
82         & -0.52952    & -0.4070786244 \\ \hline
83         & -0.52952    & -0.3972980299 \\ \hline
84         & -0.52952    & -0.401852009  \\ \hline
85         & -0.52952    & -0.4013366986 \\ \hline
86         & -0.52952    & -0.4078261972 \\ \hline
87         & -0.52952    & -0.3935246526 \\ \hline
88         & -0.52952    & -0.4035197528 \\ \hline
89         & -0.52952    & -0.4009673514 \\ \hline
90         & -0.52952    & -0.4068310896 \\ \hline
91         & -0.52952    & -0.4011399496 \\ \hline
92         & -0.52952    & -0.407983523  \\ \hline
93         & -0.52952    & -0.4008302665 \\ \hline
94         & -0.52952    & -0.4089342389 \\ \hline
95         & -0.52952    & -0.404329474  \\ \hline
96         & -0.52952    & -0.4098480158 \\ \hline
97         & -0.52952    & -0.4068971478 \\ \hline
98         & -0.52952    & -0.3970568892 \\ \hline
99         & -0.52952    & -0.4050681684 \\ \hline
100        & -0.52952    & -0.4176714243 \\ \hline
\end{tabular}
\end{table}

\begin{table}[H]
\begin{tabular}{|l|l|l|}
\hline
Iterations & Theoretical & Experimental  \\ \hline
101        & -0.52952    & -0.4127125846 \\ \hline
102        & -0.52952    & -0.4028407225 \\ \hline
103        & -0.52952    & -0.397501173  \\ \hline
104        & -0.52952    & -0.4120374641 \\ \hline
105        & -0.52952    & -0.4085077118 \\ \hline
106        & -0.52952    & -0.4095649677 \\ \hline
107        & -0.52952    & -0.4097617167 \\ \hline
108        & -0.52952    & -0.4122036682 \\ \hline
109        & -0.52952    & -0.4182641555 \\ \hline
\end{tabular}
\end{table}



\end{document}